\documentclass[journal]{IEEEtran}
\usepackage{amsmath,amssymb,amsfonts}
\usepackage{algorithmic}
\usepackage{algorithm}
\usepackage{array}
\usepackage[caption=false,font=normalsize,labelfont=sf,textfont=sf]{subfig}
\usepackage{textcomp}
\usepackage{stfloats}
\usepackage{url}
\usepackage{verbatim}
\usepackage{graphicx}
\usepackage{cite}
\usepackage{soul}

\usepackage{wrapfig}
\usepackage{tikz}
\usetikzlibrary{arrows}
\usepackage{bm}

\def\BibTeX{{\rm B\kern-.05em{\sc i\kern-.025em b}\kern-.08em
		T\kern-.1667em\lower.7ex\hbox{E}\kern-.125emX}}
	
\begin{document}

\title{Adaptive Attitude Estimation Using a Hybrid Model-Learning Approach}

\author{Eran Vertzberger and Itzik Klein \IEEEmembership{Senior Member, IEEE}
\thanks{E. Vertzberger and I. Klein are with the Hatter Department of Marine Technologies, University of Haifa, Israel (e-mail: vertzeran@gmail.com; kitzik@univ.haifa.ac.il). }}

\markboth{Adaptive Attitude Estimation Using a Hybrid Model-Learning Approach, VOL. XX, NO. XX, XXXX 2022}
{Shell \MakeLowercase{\textit{et al.}}}

%\IEEEpubid{0000--0000/00\$00.00~\copyright~2022 IEEE} 

\maketitle

\begin{abstract}
Attitude determination using the smartphone's inertial sensors poses a major challenge due to the sensor low-performance grade and variate nature of the walking pedestrian. In this paper, data-driven techniques are employed to address that challenge. To that end, a hybrid deep learning and model based solution for attitude estimation is proposed. Here, classical model based equations are applied to form an adaptive complementary filter structure. Instead of using constant or model based adaptive weights, the accelerometer weights in each axis are determined by a unique neural network. The performance of the proposed hybrid approach is evaluated relative to popular model based approaches using experimental data.
\end{abstract}


\section{Introduction}
Micro-electro-mechanical systems (MEMS) inertial sensors have revolutionized the consumer market. This is due to the availability of advanced technology enabling accurate sensing and with respect to size, weight, and power usage, it is light and small. Inertial sensors are commonly integrated in many modern products such as smartphones, autonomous drones, virtual reality headsets, and low cost robotic toys.\\
An inertial measurement unit (IMU) contains three orthogonal gyroscopes, accelerometers, and, frequently, magnetometers. IMU is used to measure angular velocity, specific force, and magnetic field, mainly for navigation purposes. A common navigation task is orientation measurement, often referred to as an attitude and heading reference system (AHRS) \cite{titterton2004strapdown}. A popular definition of attitude in the literature is the description of the relative angular displacement between two reference frames \cite{farrell2008aided}. Attitude determination is a basic requirement in many disciplines and platforms such as motion tracking \cite{madgwick2011estimation, hyde2008estimation,fourati2010nonlinear}, pedestrian dead reckoning \cite{diaz2014standalone, diaz2015evaluation, michel2017attitude, renaudin2014magnetic}, and aerial vehicles tracking \cite{mahony2008nonlinear, pourtakdoust2007adaptive,de2011uav,wang2015quaternion}. Attitude can also be estimated based on images from an on-board camera \cite{rehbinder2003pose,lobo2003vision}. Although very effective, image processing relies on the quality of the image, which is in many cases not sufficient or even worst, unavailable (if the camera is placed in a bag). Also image processing requires heavy calculation resources and is available as frequently as the cameras capture rate. IMU based AHRS is an accurate, reliable, and a computationally efficient task that can be carried out at a fast sensor rate. \\
The main challenge in AHRS is to deal with external accelerations and magnetic disturbances, such as determining smartphone attitude while a pedestrian is accelerating. The common smartphone integrates low-cost MEMS IMU readings, which are characterized by low resolution signals incorporating high amplitude noise and time varying bias. To achieve an accurate attitude solution, the nine dimensional IMU signals must be fused properly. The fusion algorithm must consider each sensor's signal composition for it to be weighted correctly in the attitude solution. Most algorithms assume a constant signal composition (equal for all axes), but, as discussed later, the amplitude of disturbances incorporated in the sensor signal varies based on time, dynamics, and location. A few classical studies addressed this issue in the literature by using assumptions and heuristics as a basis for a weighting policy. Although these methods have achieved some success, a more appealing strategy is to use data-driven approaches. In the last decade, ground breaking results have been obtained with learning approaches in fields such as image processing \cite{andersson2019deep}, speech recognition\cite{vaswani2017attention}, and even navigation \cite{asraf2021pdrnet}. Adaptation of learning approaches for attitude estimation is a new trend, making this field of study intriguing for many researchers. \\
This paper rests on the foundations of our prior studies \cite{vertzberger2021attitude,vertzberger2021Magnetometers}, addressing the integration of a model-based and an optimization-based data driven method for the purpose of attitude adaptive determination using low-cost inertial sensors. Based on this approach, a hybrid model-learning framework for attitude estimation is proposed. The learning algorithm aims to cope with the varying amplitudes of linear accelerations applied by the walking user. The output of a neural network (NN) is plugged into model equations to form an adaptive complementary filter structure. That is, instead of using constant or model-based adaptive weights, the accelerometer weights in each axis are determined by a unique NN. Instead of end-to-end deep learning solutions using a "black box" strategy (input inertial data, output attitude), the proposed method relies on a model and uses learning to determine only the required weights within the model equations. The motivation of such s strategy is to make use of the established knowledge of the deterministic nature of this problem on the one hand, with stochastic behavior learned from data on the other.\\
To demonstrate the improvement in performance, the new approach is compared to \cite{vertzberger2021attitude} and also to popular AHRS such as \cite{madgwick2011estimation,mahony2008nonlinear,farrell2008aided}. Our analysis uses experimental data collected in \cite{yan2018ridi} for four different smartphone locations: 1) Pocket, 2) Texting, 3) Body, and 4) Bag.\\
The rest of the paper is organized as follows: Section \ref{Section_LiteratureReview} gives a thorough review of the various available in the literature, which form the basis for this research. Section \ref{Section_Problem Formulation} describes the mathematical background needed for the derivation of the proposed approach. Section \ref{Section_ProposedApproach} presents the proposed framework; namely, the structure of the adaptive filter, the proposed data-driven structure, and the training process of the model. A comparison of the proposed approach to other AHRS approaches is in Section \ref{Section_AnalysisAndResults}, Section \ref{Section_Conclusions} gives the conclusions.
\section{Literature Review}	\label{Section_LiteratureReview}
Attitude estimation has been studied widely in the literature for many years. The Triad method proposed by \cite{black1964passive} is one of the earliest solutions to attitude determination. Two-directional measurements, namely magnetic field and a 3D solar cell, are considered to compute the attitude rotation matrix of a satellite. Wahba \cite{wahba1965least} formulated this problem to calculate the optimal orthogonal matrix given two or more directional measurements by optimizing a least squares cost function. The QUaternion ESTimator (QUEST) \cite{shuster1981three} method calculates an optimal attitude solution in a quaternion form. In \cite{bar1986frequency}, the authors proposed a complementary filter with an Euler angles representation of the attitude. Gyroscope and accelerometer data are processed to estimate the pitch and roll angles. A frequency domain analysis is proposed as a method for the tuning procedure. A thorough historical review of the early stages of attitude determination is given in \cite{valenti2015keeping}.\\
Due to the nonlinear relations between the vector observations and attitude representation, the extended Kalman filter (EKF) is probably the most accepted approach in the literature. In this framework, the attitude kinematic equations,  numerically propagated by gyro measurements, are treated as the system model. The accelerometer and magnetometer vector observations are processed with a linearization of the observation model to update the estimation and propagate the covariance matrix. An optimal solution is calculated by minimizing the error covariance based on assumptions on the distribution and magnitude of the process and measurement noise covariance. The EKF can be divided into two categories: a total state and an error state implementations. In the total state approach, employed in \cite{renaudin2014magnetic,sabatini2011kalman,munguia2011attitude}, a vector  quaternion representation of the attitude is estimated. In the error state approach, a state vector consisting of misalignment angles is promoted to update attitude represented in the orthogonal matrix form. Examples of error state EKF can be found in \cite{farrell2008aided} and \cite{park2020adaptive}. The generalization of this problem as a Kalman Filtering (KF) estimation scenario enables the use of state augmentation. The gyro bias and linear accelerations are modeled and added to the estimation problem to improve accuracy. Furthermore, instead of assuming a zero mean and a normal distribution, in the unscented Kalman filter (UKF), deterministic sampling is used to approximate the mean and covariance of the state distribution. This framework was used for attitude determination in \cite{diaz2014standalone,de2011uav}.\\
The work described in \cite{choukroun2006novel} creatively formulates AHRS as a linear system. This can be done by constructing of the system process matrix using the gyro measurement; this step is common when using a quaternion form. Further, the observation matrix is constructed using the accelerometer and magnetometer readings. Another derivation of a KF approach to AHRS is proposed in \cite{lee2012estimation}. A unique selection of the state vector leads to linear process and measurement models. The simplified approach in \cite{wang2015quaternion} and\cite{feng2017new} converts the gravity measurement to an observation of the attitude quaternion.\\
Although KF based solutions are well established mathematically, they are known to be sensitive to the tuning of process and measurement covariance matrices. Another drawback is the computation complexity involved in calculating the filter gain. Complementary filters (CF) have the advantage of an easier and intuitive tuning process and demand less computing resources. Examples of CF are given in \cite{bar1986frequency,mahony2008nonlinear,fourati2010nonlinear,madgwick2011estimation}. In \cite{bar1986frequency}, body rates are converted to predictions of the pitch and roll angles derivatives.  The gravity measurement and prediction are then subtracted to produce a gravity residual. The residual is filtered and used to produce updated estimations of pitch and roll derivatives. In \cite{mahony2008nonlinear,fourati2010nonlinear}, the accelerometers and magnetometers are processed to produce an estimation of the attitude quaternion error. The attitude error quaternion is used to update a predicted estimation of the quaternion derivative. A very popular algorithm is the Madgwick \cite{madgwick2011estimation} filter. The Jacobeans of  gravity and magnetic measurement models are used for a gradient descent step. The step length is the tunable parameter and constitutes the filter gain.\\
Performance of all solutions mentioned above depends on parameter tuning that reflects  the characteristics of the scenario (such as each sensor's noise and amplitude of disturbances). It is well known that the tuning procedure of a filter to a specific environment is not a trivial task and that the achieved performance degrades if the environment is not maintained. Moreover, in a dynamic environment, static parameters does not lead to optimal performance. This brings a motivation for an algorithm that tunes the fusion weights online to reflect  variations in the environment. A few studies in the literature address the AHRS problem in an adaptive Kalman filter (AKF) approach. Such an algorithm is proposed in \cite{choukroun2006novel}. Noise covariance of measurement and process are adjusted to achieve consistency with the measured residual. This is done by formulating an optimization problem to which an analytic solution is available. In \cite{pourtakdoust2007adaptive}, the measured residual is used to adjust a recursive least square model with exponential age weighting. The resulting model is used to tune the measurement noise covariance matrix. In \cite{park2020adaptive}, a minimum volume covering ellipsoid algorithm is used to adjust the measurement covariance, while in \cite{ding2007improving}, the process noise covariance is adjusted based on the measured segment of the residual samples. An interesting algorithm is presented in \cite{wang2004adaptive}, where the measurement noise is adjusted proportionally to the measured acceleration norm. Additionally, an adjustable scale factor is used to control the responsiveness of the adaptation process.\\
It is important to stress that the adaptation of noise and measurement covariance of a KF relies on the validity of the filter assumptions. In the derivation of an AHRS problem as in a KF framework, linear accelerations and magnetic disturbances are treated as Gaussian distributed uncorrelated signals. Yet, neither of the mentioned signals meet these definitions. %but are rather deterministic signals dependent on the user dynamics and projections of local magnetic fields sources. 
The violation of these underlining assumptions lead to a non-optimal performance of the AKF compared to the non-adaptive filters. Nevertheless, the value in robustness and self-tunable adaptive algorithms is significant in some applications. This was shown in \cite{vertzberger2021attitude} where \cite{pourtakdoust2007adaptive,ding2007improving} were compared to a few non-adaptive algorithms optimized for optimal performance on the tested data.\\ 
Several other papers propose integration of heuristics to deal with variations in the environment. Examples can be found in \cite{diaz2015evaluation,diaz2014standalone,michel2017attitude,renaudin2014magnetic}. The vector observation norms are compared to known values of gravity and magnetic fields. This operation is performed to detect the presence of disturbances in the sensor measurements, which would then lead to rejection of the disturbed measurements. Disturbance detection can be performed for each sample \cite{diaz2015evaluation,diaz2014standalone} or based on a buffer of measurements like in \cite{michel2017attitude,renaudin2014magnetic}.\\
The task of choosing the best among the multitude of model based AHRS algorithms is not an easy one. Informative performance surveys are given in \cite{michel2017attitude,diaz2015evaluation,vertzberger2021attitude}. 
AHRS determination of a smartphone is characterized by low-cost sensors and a dynamic/magnetic environment of variate nature. The inherent deficiencies of the classic approaches and the availability of recorded data constitute a fertile substrate to a new era of data driven solutions. Recently a few studies featured neural networks (NN) for the purpose of attitude estimation. When approaching the task of developing a data driven attitude estimator, there is no agreed convention for structure, data handling, training process, network type, input, etc. In \cite{al2019deep,russo2020danae,russo2021danae++} a deep neural network (DNN) is used to enhance the performance of the state estimator. In \cite{brossard2020denoising} a convolutional neural network (CNN) is used to denoise the gyroscope measurements of an IMU by estimating calibration parameters and bias. This process substantially improved the accuracy of attitude estimation calculated by gyro integration. In \cite{esfahani2019orinet} a long short term memory (LSTM) based architecture called OriNet was proposed to propagate the attitude estimation using the gyro measurements. The proposed structure consists of a genetic algorithm for gyro bias estimation. This network could also be thought of as a calibration process to use with a single IMU  on which it was trained. A very distinguishable approach is presented in \cite{weber2005neural,weber2021riann}, employing a pure end-to-end strategy in which the NN is fed by the raw sensor data and delivers prediction of the attitude quaternion, was employed. State-of-the-art DL tools are used to achieve the optimal choice of network configuration and hyper parameter tuning. Several publicly available datasets are used for training, achieving a robust solution.\\ 
In \cite{chen2018ionet,yan2019ronin,yan2018ridi,andersson2019deep} DNNs were used for pedestrian dead reckoning (PDR). Although not the topic of this work, the use of DNNs to process IMU data for navigation purposes is relevant. In \cite{chen2018ionet} a bi-directional LSTM was used on segments of IMU samples to predict the change in heading and location. In \cite{yan2019ronin} a ResNet \cite{he2016deep} structure was used to regress a velocity vector in the horizontal plane. In \cite{yan2018ridi} same task was carried out by support vector regression\cite{smola2004tutorial} and a support vector machine.\\
In \cite{vertzberger2021attitude,vertzberger2021Magnetometers}, a model-data-driven approach to AHRS was proposed by the authors. The main concept was to use the well established mathematical relations of attitude and vector observations but to replace tuning, heuristics models, and assumptions with data driven decisions. The data based empirical functions, designed to predict the optimal fusion weights, are nested in a complementary filter based structure. In \cite{vertzberger2021attitude}, accelerometer weights were adjusted based on an estimation of the linear acceleration. Accelerometers and gyroscopes were used to estimate the attitude (roll and pitch angles). Estimated linear accelerations were mapped to fusion weights by the use of an empirical function. This function was obtained by an optimization process using ground truth (GT) dataset. In the following study \cite{vertzberger2021Magnetometers}, the previous fundamental concept was employed and elaborated to include magnetometer readings achieving accurate heading estimation in the presence of magnetic disturbances. Additionally the heading update is decoupled from the attitude update. As a result, the gravity estimation accuracy is preserved even if the heading estimation is erroneous due to a magnetic disturbed environment. This method of applying a data driven function combined with classical equations demonstrated best performance compared to a selection of common approaches from the literature. 
\section{Problem Formulation} \label{Section_Problem Formulation}
\begin{figure} [h]
	\centering
	\includegraphics[scale=0.45]{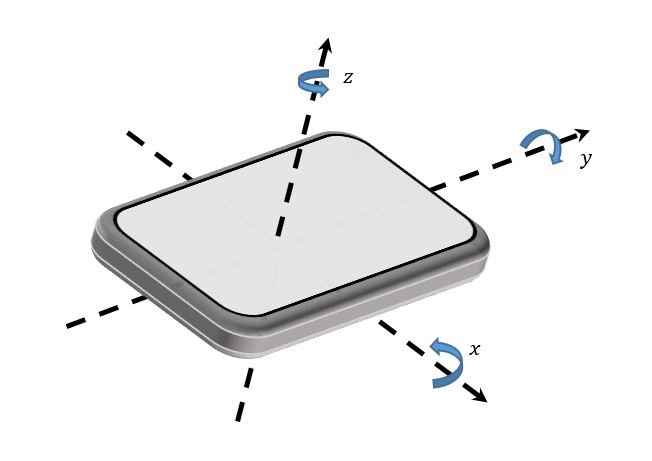}
	\caption{body reference frame fixed to a smartphone and angular rate directions.}\label{SmartphoneFrame}
\end{figure}

The attitude (roll and pitch angles) of a smartphone relative to the ground is measured based on gyroscopes and accelerometers mounted on the smartphone. Rotation matrices and Euler angles (roll, pitch, and yaw) are the attitude representations used in this work.\\
Two reference frames are employed:
\begin{itemize}
	\item \textbf{Body frame} (b-frame): fixed to the 
	IMU axes such that the x-axis points to the right side of the
	smartphone screen, y-axis points to the top of the screen, and
	z-axis points towards the outer side of the screen, completing
	the right-handed coordinate system, as pretested in Fig. \mbox{\ref{SmartphoneFrame}}.
	\item \textbf{Reference frame} (r-frame): fixed to the ground so that the x-axis points north, z-axis points down, and y-axis completes the right-handed coordinate system pointing east.
\end{itemize}

%
%\subsection{Sensors}
%	Models of body rates and specific force are given by:
%	\begin{equation}\label{GyroModel} 
	%		\tilde{\bm{\omega}}^b=\bm{\omega^b}+\bm{b}_g+\bm{v}_g
	%	\end{equation}
%	\begin{equation}\label{AccModel}  
	%		\bm{\tilde{f}}^b=\bm{f}^b+\bm{b}_a+\bm{v}_a
	%	\end{equation}
%	where $\tilde{\bm{\omega}}$ is the measurement expressed in the body frame, $\bm{\omega^b}$ is the true angular velocity vector, $\bm{b}_g$ is a bias component, $\tilde{\bm{f}}$ is the accelerometer measurement expressed in the body frame, $\bm{f}^b$ is the true specific force vector, $\bm{b}_a$ is the bias and $\bm{v}_g$ and $\bm{v}_a$ are assumed to be zero mean, uncorrelated white Gaussian noise.  The true specific force is comprised of linear acceleration and the Earth gravity vector, hence 
%	\begin{equation}\label{AccModel}  
	%		\bm{f}^b=\bm{a}^b-\bm{g}^b
	%	\end{equation}
%	where $\bm{a}^b$ is linear acceleration expressed in the body frame.
%
\subsection{Gyro Integration}      \label{Subsection_GyroIntegration}
Most AHRS algorithms use gyro integration equations as the baseline for deriving the attitude \cite{titterton2004strapdown,farrell2008aided}. To formulate the problem, transformation matrices can be used to express the change in attitude using the body rates expressed in the body frame. The orientation of the body frame relative to the reference frame is represented by a rotation matrix $\bm{R}^r_b$. For brevity, $\bm{R}^r_b$ in $\bm{R}^{3 \times 3}$  is denoted as $\bm{R}$. Employing $\bm{R}$, a vector expressed in the b-frame can be transformed to the r-frame by 
\begin{equation}\label{VectorTransformation}  
	\bm{v}^r=\bm{R}\bm{v}^b
\end{equation}
where $\bm{v}^b$ and $\bm{v}^r$ are vectors expressed in b and r frames, respectively. The derivative of the transformation matrix $\dot{\bm{R}}$ can be obtained using the angular velocity vector $\bm{\omega}^b$, 
\begin{equation}\label{RotationMatrixDerivative}
	\dot{\bm{R}}=\bm{R}\bm{\Omega}_{\times}
\end{equation}
where 
\begin{equation}\label{CrossProductOperator}
	\bm{\Omega}_{\times}=
	\begin{bmatrix}
		0         & -\omega_z & \omega_y  \\
		\omega_z  & 0         & -\omega_x \\
		-\omega_y & \omega_x  & 0         \\
	\end{bmatrix}
\end{equation}
is the skew-symmetric form of $\bm{\omega}^b$. \\
Provided with initial conditions,  using \eqref{RotationMatrixDerivative} and based only on the gyroscope readings, the derivative of the transformation matrix is
\begin{equation}\label{RotationMatrixDiscreteDerivative}
	\hat{\dot{\bm{R}}}_k=\hat{\bm{R}}_{k-1}\tilde{\bm{\Omega}}_{\times,k}
\end{equation}
where the hat and tilde signs imply an estimated and measured values, respectively.
Using a simple Euler propagator, the updated transformation is
\begin{equation}\label{DiscreteGyroIntegration}
	\hat{\bm{R}}_{g,k}=\hat{\bm{R}}_{k-1}+\hat{\dot{\bm{R}}}_k dt
\end{equation}
where $dt$ is the sampling interval, k is the time index, and the subscript $g$ indicates a gyro based estimation. Due to numerical errors originated from the discretization of the integral operation, the resulting $\hat{\bm{R}}_{g,k}$ matrix is not orthogonal \cite{farrell2008aided,titterton2004strapdown}. To treat this issue, an orthogonalization step is carried out each iteration
\begin{equation}\label{Orthogonalization}
	\hat{\bm{R}}_{g,k}^o=\hat{\bm{R}}_{g,k}(\hat{\bm{R}}_{g,k}^T \hat{\bm{R}}_{g,k})^{-\frac{1}{2}}
\end{equation} 
 The attitude then can be directly obtained from \eqref{Orthogonalization}. For simplicity the superscript $o$ will be omitted in the following use of $\hat{\bm{R}}_{g,k}^o$.

\subsection{Accelerometer Update}
Gyroscope measurements lack the information of the absolute attitude value. Therefore, when applying the numerical gyro integration procedure expressed in the previous section, $\hat{\bm{R}}_{g,k}$ will have an accumulated error factor that grows boundlessly due to the noisy nature of the gyroscope. Specific force measurements, on the other hand, possess information of the absolute angular position of the IMU (relative to Earth's gravity vector) and can be used to correct the gyro integration solution. Accelerometer updates can be carried out in several ways. In some cases, like in \cite{mahony2008nonlinear,bar1986frequency} the body rates are updated. In others, like in \cite{munguia2011attitude,sabatini2011kalman,renaudin2014magnetic}, gravity errors are  multiplied by the Kalman gain, $\bm{K}_k$, for posterior update of the attitude, incorporated in the state vector

\section{Proposed Approach}\label{Section_ProposedApproach}
The foundations for the proposed data-driven approach are based on our previous work \cite{vertzberger2021attitude,vertzberger2021Magnetometers}. There, the update weights of the accelerometers and magnetometers were determined by an empiric function obtained by an optimization process. In this paper, this empiric function is replaced by a NN model aiming to predict the optimal fusion weights based on the momentary residual. In this manner, instead of an end-to-end deep learning approach, the proposed method employs a solid theoretical AHRS algorithm and incorporates a NN to predict the weights in a hybrid fashion. 
Although a single block in the pipeline is replaced, major advantages are achieved. The NN is a differentiable function and thus enables analytical calculations of the gradient in the back propagation stage. The analytical calculation replaces a forward difference gradient approximation used in our former proposed approaches. Moreover, the training process depicted in Section \ref{Subsection_Training_Process} adopts batch optimization and therefore only a portion of the training set is processed in each training iteration. Another benefit is that the NN generalizes from a batch of shuffled short segments to full length experiments and thus is more robust. The framework, which is comprised of the structure paired with the proposed training process is not only superior in performance but is also much faster to optimize. \\ 
The proposed model focuses on gravity estimation (like in \cite{vertzberger2021attitude}) without the use of magnetometers. Nevertheless the same solution can also be used for heading estimation (like in \cite{vertzberger2021Magnetometers}).\\ 
The proposed model consists of two supplementary components: a complementary model-based filter structure and a NN for gain determination that requires a training procedure. We highlight two properties that enable accurate attitude estimation in a dynamic environment:
\begin{enumerate}
	\item \textbf{Deep Learning (DL) Based Adaptation Policy} \textendash The linear accelerations are estimated and inserted to a NN. The NN is trained, as described in Section \ref{Subsection_Training_Process}, to determine the update weights of the filter for optimal performance on the train set. 
	\item \textbf{Component Based Update} \textendash The gravity is updated in each axis separately based on the estimated acceleration in the appropriate direction (Section \ref{Subsection_UpdateStage}). 
\end{enumerate}
\subsection{Overview}
The structure of the proposed approach, denoted as deep attitude estimator (DAE), is presented in Fig. \ref{FigureOfFilterStructure}. In the figure, the kinematic equations block contains the gyro integration process described in Section \ref{Subsection_GyroIntegration}. The transformation to the b-frame is carried out using  (\ref{GravityProjectionAfterGyroIntegration}). The left branch of the diagram symbolizes the adaptive process described in Section \ref{Subsection_AdaptiveFilterGainDetermination}. The "Update" and "Triad" blocks denote the update stage explained in Section \ref{Subsection_UpdateStage}.
\tikzstyle{block} = [rectangle, fill=white, rounded corners,
minimum height=1cm, minimum width=1cm,text centered,text width=2cm,draw=black]
\tikzstyle{sum} = [circle, fill=white, draw=black]
\tikzstyle{io} = [coordinate]
\tikzstyle{arrow} = [thick,->,>=stealth]
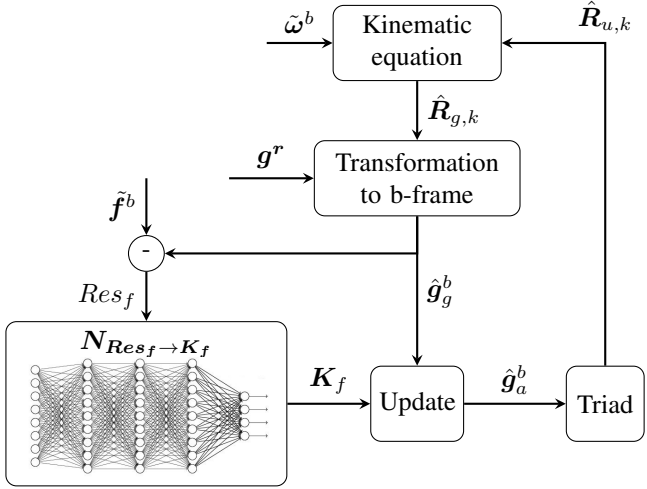
\begin{figure} [h!]
	\centering	
	\begin{tikzpicture}[auto, node distance=1.5cm,>=latex']
		\node [io, name=GyroInput] (GyroInput) {Gyro};
		\node [block, right of=GyroInput, xshift=0.5cm] (KinematicEquationBlock) {Kinematic equation};
		\node [block, below of=KinematicEquationBlock, yshift=-0.3cm, text width=2.5cm] (R2gb){Transformation to b-frame};
		\node [io, left of=R2gb, xshift=-1cm, name=gn] (gr) {};
		\node [block, below of=R2gb, yshift=-1.5cm ,text width=1cm ] (UpdateBlock) {Update};  % \ref{VectorTransformation}
		\node [block, left of=UpdateBlock, xshift=-2.1cm ,text width=3.5cm] (GainMapBlock) {$\bm{N}_{\bm{Res_f}\rightarrow \bm{K_f}}$ \includegraphics[width=1.\textwidth]{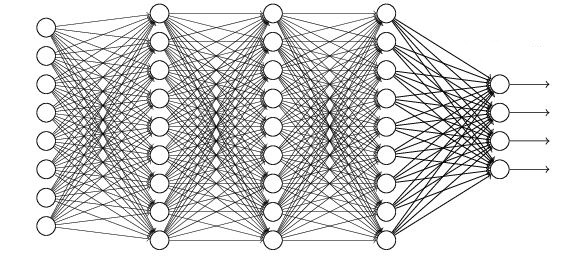}};
		\node [block, right of=UpdateBlock, xshift=1.cm,text width=0.8cm] (TriadBlock) {Triad};
		\node [sum,   above of=GainMapBlock, yshift=0.5cm] (sum) {-};
		\node [io, above of=sum, yshift=-0.5cm, name=AccInput] (AccInput) {};
		\draw [arrow] (GyroInput) --node[anchor=south] {$\tilde{\bm{\omega}}^b$} (KinematicEquationBlock);
		\draw [arrow] (KinematicEquationBlock) --node {$\hat{\bm{R}}_{g,k}$} (R2gb);
		\draw [arrow] (R2gb) --node {$\hat{\bm{g}}^b_{g}$} (UpdateBlock);
		\draw [arrow] (UpdateBlock) -- node[anchor=south] {$\hat{\bm{g}}_a^b$}(TriadBlock);
		\draw [arrow] (sum) --node[anchor=east] {$Res_f$} (GainMapBlock);
		\draw [arrow] (GainMapBlock) --node[anchor=south] {$\bm{K}_f$} (UpdateBlock);
		\draw [arrow] (AccInput) -- node[anchor=east] {$\tilde{\bm{f}}^b$} (sum);
		\draw [arrow] (R2gb) |- (sum);
		\draw [arrow] (TriadBlock) |- node[anchor=south] {$\hat{\bm{R}}_{u,k}$} (KinematicEquationBlock);
		\draw [arrow] (gr) --node[anchor=south] {$\bm{g^r}$} (R2gb);
	\end{tikzpicture}
	\caption{System flowchart} \label{FigureOfFilterStructure}
\end{figure}
\subsection{Component Based Update}\label{Subsection_UpdateStage}
In general, linear accelerations cause the dominant error in the attitude estimation, as most AHRS algorithms assume zero linear acceleration during their operation. Thus, in the presence of disturbances (linear accelerations), small weights are required to ignore the accelerometer measurements. Yet, care must be taken as the linear accelerations are not equal in all axes, thus different weights should be applied. To circumvent this deficiency, the state update is expressed using an estimate of the gravity expressed in the b-frame.
\begin{equation}\label{GravityProjectionAfterGyroIntegration} 
	\hat{\bm{g}}_g^{b}=\hat{\bm{R}}_{g,k}\bm{g}^{r}
\end{equation}
Finally, the component based accelerometer update is 
\begin{equation}\label{ComponentBasedAccUpdate} 
	\hat{\bm{g}}_a^{b}=\hat{\bm{g}}_g^{b}+\bm{K_f} \left(\tilde{\bm{f}}^b-\hat{\bm{g}}_g^{b} \right)
\end{equation}
where $\tilde{\bm{f}}^b$ is the specific force measurement and $\bm{g}^r$ is the known gravity field vector expressed in the  r-frame. 
Subscript $a$ stands for an accelerometer based estimate. The gain matrix is
\begin{equation}\label{AccWeightsDefinition} 
	\bm{K_f}=
	\begin{bmatrix}
		k_{fx}& 0      & 0  \\
		0     & k_{fy} & 0  \\
		0     & 0      & k_{fz}
	\end{bmatrix}
\end{equation}
and $k_{fx},k_{fy},k_{fz}$ are the accelerometer weights in each axis limited to values ranging between $0$ to $1$. \\
The updated gravity vector is used to create a transformation matrix, $\hat{\bm{R}}_{u,k}$, where the subscript $u$ denotes an updated estimation. The conversion of the gravity  vector to a transformation matrix is performed using (\ref{TriadTemporalFrameDefinitionXInReferenceFrame})-(\ref{TriadBodyToReferenceMatrix}), which are based on the Triad approach \cite{black1964passive}. This method, employed in \cite{vertzberger2021Magnetometers},  uses a magnetic field measurement. As magnetic data is not used in this work, pseudo magnetic field measurements, $\bm{m}^r$ and $\bm{m}^b$ are employed instead.    \\
A temporal frame (t-frame) is defined using the r-frame components. The $x$ component of the t-frame is defined as the direction of the gravity vector
\begin{equation}\label{TriadTemporalFrameDefinitionXInReferenceFrame} 
	\bm{x}^r_t=\bm{g}^r
\end{equation}
The $y$ component of the t-frame is defined as a direction perpendicular to the gravity vector and pseudo-magnetic field vector.
\begin{equation}\label{TriadTemporalFrameDefinitionYInReferenceFrame}
	\bm{y}^r_t=\frac{\bm{g}^r\times \bm{m}^r}{\|\bm{g}^r\times \bm{m}^r\|}
\end{equation}
where
\begin{equation}\label{PseudoMagneticFiledInReferenceFrame}
	\bm{m}^r=\begin{bmatrix} 1 & 0 & 0 \end{bmatrix}^T
\end{equation}
The t-frame $z$ component completes the right hand system
\begin{equation}\label{TriadTemporalFrameDefinitionZInReferenceFrame}
	\bm{z}^r_t=\frac{\bm{x}^r_t\times \bm{y}^r_t}{\|\bm{x}^r_t\times \bm{y}^r_t\|}
\end{equation}
Similarly, the  t-frame components can be expressed using the  b-frame components as     
\begin{equation}\label{TriadTemporalFrameDefinitionXInBodyFrame}
	\bm{x}^b_t=\hat{\bm{\bm{g}}}_a^{b}
\end{equation}
\begin{equation}\label{TriadTemporalFrameDefinitionYInBodyFrame}
	\bm{y}^b_t=\frac{\hat{\bm{\bm{g}}}_a^{b}\times \bm{m}^{b}}{\|\hat{\bm{\bm{g}}}_a^{b}\times \bm{m}^{b}\|}
\end{equation}
\begin{equation}\label{TriadTemporalFrameDefinitionZInBodyFrame}
	\bm{z}^b_t=\frac{\bm{x}^b_t\times \bm{y}^b_t}{\|\bm{x}^b_t\times \bm{y}^b_t\|}
\end{equation}
where 
\begin{equation}\label{PseudoMagneticFiledInBodyFrame}
	\bm{m}^{b}=\hat{\bm{R}}_{g,k}\bm{m}^{r}
\end{equation}
and $\bm{R}_{g,k}$ is defined in (5).\\
The t-frame components construct the columns of the transformation matrices from t-frame to r-frame using \eqref{TriadTemporalFrameDefinitionXInReferenceFrame}-\eqref{TriadTemporalFrameDefinitionZInReferenceFrame}
\begin{equation}\label{TriadTemporalToReferencerotationMatrix}
	\bm{R}^r_t=\left[\bm{x}^r_t | \bm{y}^r_t |\bm{z}^r_t\right]
\end{equation}
and from t-frame to b-frame using \eqref{TriadTemporalFrameDefinitionXInBodyFrame}-\eqref{TriadTemporalFrameDefinitionZInBodyFrame}
\begin{equation}\label{TriadTemporalToBodyMatrix}
	\bm{R}^b_t=\left[\bm{x}^b_t | \bm{y}^b_t |\bm{z}^b_t\right]
\end{equation}
Hence the updated transformation is 
\begin{equation}\label{TriadBodyToReferenceMatrix}
	\hat{\bm{R}}_{u,k}=\bm{R}^r_t\left(\bm{R}^b_t\right)^T
\end{equation}
and can be used in the next iteration as $\hat{\bm{R}}_{u,k-1}$ in \eqref{RotationMatrixDiscreteDerivative}.
\subsection{Weight Prediction}\label{Subsection_AdaptiveFilterGainDetermination}
Propagation of (\ref{RotationMatrixDiscreteDerivative})-(\ref{Orthogonalization}) is carried out using the gyroscope measurements. The resulting transformation matrix $\hat{\bm{R}}_{g,k}$ is plugged into (\ref{GravityProjectionAfterGyroIntegration}) to yield the predicted gravity vector $\hat{\bm{g}}^b_g$. Next, the measurement residuals, required in (\ref{ComponentBasedAccUpdate}) for the update stage, are obtained by  
\begin{equation}\label{AccResidualDefinition} 
	\bm{Res_f}=\tilde{\bm{f}}^b-\hat{\bm{g}}_g^{b}
\end{equation}
and mapped via
\begin{equation}\label{AccWeightsCalculation} 
	\bm{K_f}=\bm{N}_{\bm{Res_f}\rightarrow \bm{K_f}}(\bm{Res_f})
\end{equation}
to the accelerometer weights. The ``Gain Net", $\bm{N}_{\bm{Res_f}\rightarrow \bm{K_f}}$, is the NN that predicts the values of the diagonal of (\ref{AccWeightsDefinition}) based on $\bm{Res_f}$. It is obtained by performing a training process based on experimental data (sampled sensors and matching ground truth attitude values) with the objective of minimizing the attitude loss function. The attitude loss function expresses an RMS calculation of error angle between the GT and estimated gravity vectors: 
\begin{equation}\label{GravityLossFunction}    
	\underset{\bm{f}_{\bm{Res_f}\rightarrow \bm{K_f}}}{\text{min}} J_f=
	\sqrt{\frac{1}{n}\sum_{i=1}^{n}\left[\cos^{-1}\left(\bm{g}^b_{GT,i}\cdot\hat{\bm{g}}^b_{a,i}\right)\right]^2}
\end{equation}
where $n$ is the number of samples in a segment of an experiment and $\bm{g}^b_{GT,i}$ is the gravity GT values. 
\subsection{Network Structure} \label{Subsection_Network_Structure}
\begin{figure} [h!]
	\centering
	\includegraphics[scale=0.3]{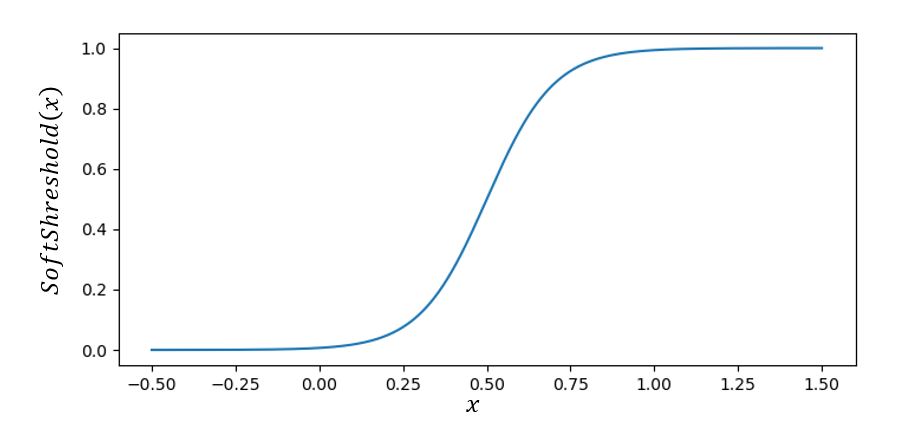}
	\caption{Soft threshold function used in the output layer.}\label{SoftShresholdFigure}
\end{figure}

\begin{figure} [h!]
	\centering	
	\begin{tikzpicture}[auto, node distance=1.5cm,>=latex']
		\node [io](Input) {};
		\node [block, right of=Input, xshift=0cm,rotate=90,transform shape] (AugmentationBlock) {Input \\ Augmentation};
		\draw [arrow] (Input) --node[anchor=south] {$Res_{f_i}$} (AugmentationBlock);
		
		\node [block, right of=AugmentationBlock, xshift=-0.2cm,text width=2.3cm,minimum height=0.5cm, rotate=90,transform shape] (L1) {\tiny$[9\times16]$};
		\node [block, right of=L1, xshift=-0.97cm,text width=2.3cm,minimum height=0.5cm, rotate=90,transform shape] (Tanh1) {tanh};
		\draw [arrow] (AugmentationBlock) --node[anchor=south] {\tiny$[9]$} (L1);
		
		\node [block, right of=Tanh1, xshift=-0.5cm,text width=2.3cm,minimum height=0.5cm, rotate=90,transform shape] (L2) {\tiny$[16\times32]$};
		\node [block, right of=L2, xshift=-0.97cm,text width=2.3cm,minimum height=0.5cm, rotate=90,transform shape] (Tanh2) {tanh};
		\draw [arrow] (Tanh1) --node[anchor=south] {} (L2);
		
		\node [block, right of=Tanh2, xshift=-0.5cm,text width=2.3cm,minimum height=0.5cm, rotate=90,transform shape] (L3) {\tiny$[32\times64]$};
		\node [block, right of=L3, xshift=-0.97cm,text width=2.3cm,minimum height=0.5cm, rotate=90,transform shape] (Tanh3) {tanh};
		\draw [arrow] (Tanh2) --node[anchor=south] {} (L3);
		
		\node [block, below of=Tanh3, yshift=-1.5cm,text width=2.3cm,minimum height=0.5cm, rotate=90,transform shape] (L4) {\tiny$[64\times32]$};
		\node [block, left of=L4, xshift=0.97cm,text width=2.3cm,minimum height=0.5cm, rotate=90,transform shape] (Tanh4) {tanh};
		\draw [arrow] (Tanh3) --node[anchor=south] {} (L4);
		
		\node [block, left of=Tanh4, xshift=0.5cm,text width=2.3cm,minimum height=0.5cm, rotate=90,transform shape] (L5) {\tiny$[32\times1]$};
		\node [block, left of=L5, xshift=0.18cm,text width=2.35cm,minimum height=0.5cm, rotate=90,transform shape] (SoftThreshold) {\includegraphics[width=1.0\textwidth]{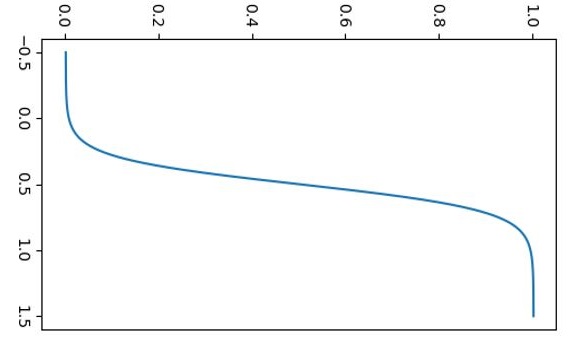} \\ \tiny$Soft \  Threshold$ };
		\draw [arrow] (Tanh4) --node[anchor=south] {} (L5);

		\node [io, left of=SoftThreshold, xshift=-0.5cm](Output) {};
		\draw [arrow] (SoftThreshold) --node[anchor=south] {$K_{f_i}$} (Output);
	\end{tikzpicture}
	\caption{Network structure for the DAE framework.} \label{NetworkStructureBlockDiagram}
\end{figure}
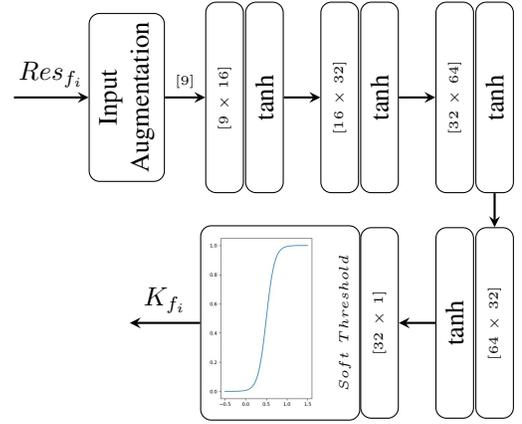

There are more than a few possibilities for network configuration. The chosen structure consists of three single-input-single-output NN with identical hyper parameters. Each is fed by an element of the acceleration residual calculated in (\ref{AccResidualDefinition}) and predicts each axis acceleration weight in (\ref{AccWeightsDefinition}). To encourage nonlinear behavior, the scalar input to the NN is augmented to a vector containing powers of the scalar input
\begin{equation}\label{NetworkInputAugmentation} 
	\bm{u}_i=\begin{bmatrix} Res_{f_i}^{p},& ... , &1,& Res_{f_i},& ... ,&Res_{f_i}^{P}  \end{bmatrix}^T
\end{equation}
where $u$ is the NN input in a certain axis with a subscript $i = [x,y,z]$, $p$ is the smallest power, and $P$ is the highest power. The values chosen for smallest and highest power are $p=-3$ and $P=5$, resulting in an input dimension of $\bm{u}_i\in\mathtt{R}^{9}$. We limit $Res_{f_i}$ to a minimal value of $1\times10^{-4}$ to avoid big numbers with negative powers. The NN structure in each axis is denoted by $NN_i(u_i)$ and comprises five dense layers. The chosen layer sizes are [16, 32, 64, 32, 1]. The layers are connected by $tanh()$ activation functions except for the output layer. Since the NN output is substituted to the multiplicative gain matrix $K_f$, the values must be in the range of [0,1]. Therefore,  a soft threshold function was defined as the activation function of the output layer
\begin{equation}\label{SoftThresholdDefinition}
	SoftThreshold(x)=tanh((x-0.5)\times5)\times0.5+0.5
\end{equation}
and the gain matrix is
\begin{equation}\label{SoftThresholdUse}
	K_{f_i} = SoftThreshold(NN_i(u_i))
\end{equation}
As shown in Fig.~\ref{SoftShresholdFigure}, the soft threshold function has a continuous derivative and thus is more appropriate than simple thresholds for a deep learning training process. 
The overall structure of the network is depicted in Fig.~\ref{NetworkStructureBlockDiagram} 
\subsection{Data Handling and Training Process}\label{Subsection_Training_Process}
\begin{figure} [h!]
	\centering	
	\begin{tikzpicture}[auto, node distance=1.5cm,>=latex']
		\node [io](SensorsInput) {};
		\node [block, right of=SensorsInput, xshift=1cm] (ClassicEquationsBlock) {CF Classic Equations};
		\node [io, above of=ClassicEquationsBlock, yshift=-0.3cm](InitialCondition) {};
		\draw [arrow] (SensorsInput) --node[anchor=south] {Sensors} (ClassicEquationsBlock);
		\draw [arrow] (InitialCondition) --node[anchor=east] {$\hat{\bm{R}}_{g,k-1}$} (ClassicEquationsBlock);
		\node [block, right of=ClassicEquationsBlock, xshift=2.0cm] (NN Block) {$\bm{N}_{\bm{Res_f}\rightarrow \bm{K_f}}$};
		\node [io, below of=ClassicEquationsBlock, yshift=+0.3cm](output) {};
		\draw [arrow] (ClassicEquationsBlock) --node[anchor=east] {$\hat{\bm{R}}_{g,k}$} (output);
		\path[->,thick] 
		([yshift=-2mm] ClassicEquationsBlock.north east) edge node {$Res_f$} ([yshift=-2mm] NN Block.north west)
		([yshift=+2mm] NN Block.south west) edge node {$K_f$} ([yshift=+2mm] ClassicEquationsBlock.south east)
		;
	\end{tikzpicture}
	\caption{DAE is represented as in an RNN structure.} \label{FilterAsRnn}
\end{figure}
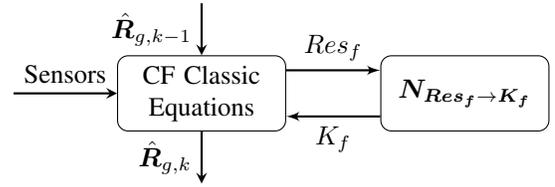

To better explain the process, the structure of the solution is rearranged and presented in Fig.~\ref{FilterAsRnn}. The right side of Fig.~\ref{FigureOfFilterStructure}, which contains the model-based equations of the complementary filter, is combined with the left block in Fig. \ref{FilterAsRnn} containing the NN, $\bm{N}_{\bm{Res_f}\rightarrow \bm{K_f}}$. Viewing the filter in this manner leads to the understanding that this structure actually acts like a recurrent neural network (RNN). 
Every inference of the filter requires (like a common model-based filter) a sensor sample and an initial condition, $\hat{\bm{R}}_{g,k-1}$. Therefore, to calculate the loss function, (\ref{GravityLossFunction}), for a sequence of sensor samples, requires an inference of the filter for each sensor sample with an internal state, $\hat{\bm{R}}_{g,k}$, passed in each inference of the filter. \\
The training is performed using stochastic optimization. For this purpose, the data is segmented into short sequences. The sequences are shuffled and grouped in batches with an incentive of a homogeneous distribution of the training data. The initial condition for each segment is taken from the momentary GT measurements. Using short segments for training, causes a decrease in the impact of errors caused by gyro bias on the overall loss. In a dynamic environment, which includes linear accelerations, short segments lead to a low amplitude weighting policy, which is not optimal for long sequences. To treat this issue, the initial condition to each segment is added with random errors from a uniform distribution.
\section{Analysis and Results}\label{Section_AnalysisAndResults}
\subsection{Preliminaries}
The experimental data used in this work was published in \cite{yan2018ridi} for PDR applications yet holds relevant information for AHRS analysis; i.e., the inertial sensor readings and attitude GT. The dataset contains 60 recordings, two minutes long each, sampled at $200$Hz. The smartphone is carried by six different people in four different locations: 
\begin{enumerate}
	\item \textbf{Texting}: The smartphone is held in the user's hand as if typing/reading a text message.
	\item \textbf{Pocket}:  User places the smartphone in his front pocket while walking.
	\item \textbf{Body}: User carries the smartphone on the waist with a strap.
	\item \textbf{Bag}: User walks while smartphone is located in a bag.
\end{enumerate}
GT trajectory data was obtained using a visual inertial odometry system. \\
%In \cite{vertzberger2021attitude}, the performance of a filter proposed by the authors was compared to a selection of estimators from the literature including Two CF \cite{madgwick2011estimation, mahony2008nonlinear}, an  error state EKF \cite{munguia2011attitude}, a total state EKF \cite{farrell2008aided}, Two Adaptive EKF \cite{wang2004adaptive,ding2007improving}, and pure gyro integration. The filter denoted as Adaptive Gravity Projection Estimator (AGPE) obtained best performance when tested on all subsets of the data set used in this study by an average of 28\%. 
%
To assess the proposed DAE approach, it is compared to the AGPE \cite{vertzberger2021attitude}, Madgwick \cite{madgwick2011estimation}, Mahony \cite{mahony2008nonlinear}, an error state EKF \cite{munguia2011attitude}, and a adaptive EKF (AEKF) \cite{ding2007improving}. For the sake of valid and fair comparison, each filter is tuned by an optimization process to minimize the attitude loss function (\ref{GravityLossFunction}). In addition, gyro bias correction is not applied in any of the filters. The dataset is divided into four subsets according to the smartphone placement: Pocket, Texting, Body, and Bag. Every subset is divided into a training set on which all algorithms parameters are optimized/trained, and a test set used for performance evaluation. That is, the training procedure was made separately for each smartphone location. The performance measure adopted in this work is
\begin{equation}\label{PerformanceMeasure}    
	e=\sqrt{e_\phi^2+e_\theta^2}
\end{equation}
where $e_\phi$ and $e_\theta$ are defined by
\begin{equation}\label{PhiError}    
	e_\phi=\sqrt{\frac{1}{n}\sum_{i=1}^{n}\left(\phi_{GT,i}-\hat{\phi_i}\right)^2}
\end{equation}
\begin{equation}\label{ThetaError}    
	e_\theta=\sqrt{\frac{1}{n}\sum_{i=1}^{n}\left(\theta_{GT,i}-\hat{\theta_i}\right)^2}
\end{equation}
and $\phi_{GT,i},\hat{\phi_i},\theta_{GT,i},\hat{\theta_i}$ are the GT and estimated roll and pitch angles, respectively. \\
\subsection{DAE Structure Evolution}\label{sec:DAE_st_ev}
\begin{figure} [h]
	\centering
	\includegraphics[scale=0.6]{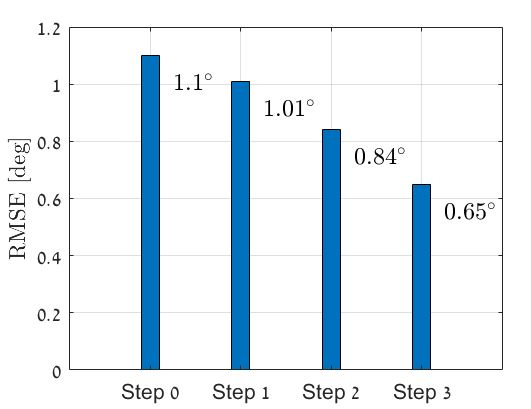}
	\caption{Performance improvement contribution achieved by each of the NN components.}\label{EvolutionFigure}
\end{figure}

The configuration of network structure and training process was achieved (like most most DL solutions) by trial and error. This section presents the workflow and emphasizes the contribution of each component to the overall performance. 
To reach the conceptual structure, experiments are performed with a diminished data set comprised of a few experiments with mixed modes. The improvement in accuracy of each development step is depicted in Fig.~\ref{EvolutionFigure} in terms of the performance measure defined in \eqref{PerformanceMeasure}.
\begin{itemize}	
	\item[$\bullet$] \textbf{Step 0}: First attempts to optimize a pure, fully connected NN using short segments ended with a performance measure of $e = 1.1$deg.
	\item[$\bullet$] \textbf{Step 1}: The resulting over-fitting solution leaned mainly on the gyro integration process (Section \ref{Subsection_GyroIntegration}) with almost no accelerometer update. The reason for this is a precise initial condition and short segments duration. Therefore, the gyro solution was accurate on the train data but, as expected, performed poorly on the long sequences of the test data. Adding errors to the initial condition forced the use of accelerometers and improved the result to $e = 1.01$deg.
	\item[$\bullet$] \textbf{Step 2}: Experience gained when developing the AGPE gave an important insight: the desired function is highly non-linear. Therefore, we decided to augment the scalar input to a polynomial, and by that promote  nonlinear behavior of the function. Input augmentation immediately improved with $e = 0.84$deg.
	\item[$\bullet$] \textbf{Step 3}: Shuffling and grouping the segments into batches, gave a homogeneous distribution of the data. As a result, the gradient converged in a smoother fashion and generalized better to the test data. This step improved the result to $e = 0.65$deg. 
\end{itemize}
\subsection{Framework Parameters}
After achieving the framework and optimizing the hyper parameters (segment size, batch size, errors amplitude, etc.) on the whole dataset we achieved the following configuration. The data is segmented into $8000$  samples of long sequences with a batch size of five samples. Based on our analysis in Section  \ref{sec:DAE_st_ev}, errors added to the GT initial condition are taken from a uniform distribution with a maximal value of $0.1$ degree. The NN was trained with an SGD optimizer with a learning rate of $3 \times 10^{-4}$. All network parameters were given in Section  \ref{Subsection_Network_Structure}. To validate the performance on long sequences, the test dataset comprises complete two-minutes experiments.
\subsection{Results}
Table~\ref{PocketTable} gives the results for the Pocket mode. The proposed approach achieved the best performance both for the roll and pitch errors and with an overall improvement of 28\% of the second best approach, AGPE.   
\setlength{\arrayrulewidth}{1pt}
\begin{table}[h!]
	\caption{Pocket mode}\label{PocketTable}
	\centering
	\begin{tabular}{| l || c | c |c |}
		\hline
		Algorithm 			 & $e_{\phi} [deg]$	& $e_{\theta} [deg]$	& $e [deg]$   \\
		\hline\hline
		Madgwick  		 	 & 0.98 		& 0.62 			& 1.16\\ 
		\hline
		Mahony    			 & 1.29 		& 0.80 			& 1.51\\
		\hline
		ES EKF    			 & 0.86 		& 0.70 			& 1.11\\
		\hline
%		AEKF 1    			 & 1.73 		& 1.5  			& 2.29\\
%		\hline
		AEKF     			 & 1.27 		& 0.83 			& 1.52\\
		\hline
		AGPE		     	 & 0.76 		& 0.56 			& 0.94\\
		\hline
		DAE (ours)     		 & $\bm{0.52}$ 	& $\bm{0.44}$ 	& $\bm{0.68}$\\
		\hline
		
	\end{tabular}
\end{table}
In Table~\ref{TextingTable} the results for the Texting mode are presented. The, overall performance of DAE is equivalent to AGPE. The Mahony filter, which is the best performing classic algorithm on this subset, had better accuracy in the pitch direction. Nevertheless, the overall performance measure of DAE is 19\% better.   
\begin{table}[h!]
	\caption{Texting mode}\label{TextingTable}
	\centering
	\begin{tabular}{| l || c | c |c |}
		\hline
		Algorithm 			 & $e_{\phi} [deg]$	& $e_{\theta} [deg]$	& $e [deg]$   \\
		\hline\hline
		Madgwick  			 & 1.06 		& 0.55 			& 1.19\\ 
		\hline
		Mahony    			 & 0.97 		& $\bm{0.41}$	& 1.05\\
		\hline
		ES EKF    			 & 1.08 		& 0.53			& 1.20\\
		\hline
%		AEKF 1    		  	 & 1.98 		& 1.11  		& 2.27\\
%		\hline
		AEKF     			 & 1.03 		& 0.63 			& 1.21 \\
		\hline
		AGPE		     	 & 0.74 		& 0.43      	& 0.86\\
		\hline
		DAE (ours)     		 & $\bm{0.68}$	& 0.51 			& $\bm{0.85}$\\
		\hline
	\end{tabular}
\end{table}
In Body mode (Table \ref{BodyTable}) AGPE performed better than DAE by 15\%. The overall performance measure of DAE is 37\% better than the Mahony filter. 
\begin{table}[h!]
	\caption{Body mode}\label{BodyTable}
	\centering
	\begin{tabular}{| l || c | c |c |}
		\hline
		Algorithm 			 & $e_{\phi} [deg]$	& $e_{\theta} [deg]$	& $e [deg]$   \\
		\hline\hline
		Madgwick  			 & 1.14 		& 0.92 			& 1.46\\ 
		\hline
		Mahony    			 & 0.86 		& 0.65			& 1.08\\
		\hline
		ES EKF    			 & 1.11 		& 0.63			& 1.28\\
		\hline
%		AEKF 1    			 & 2.45 		& 1.75  		& 3.01\\
%		\hline
		AEKF     	 		 & 1.01 		& 0.87 			& 1.33 \\
		\hline
		AGPE		     	 & $\bm{0.48}$  & $\bm{0.33}$	& $\bm{0.58}$\\
		\hline
		DAE (ours)     		 & 0.56 		& 0.44 			& 0.68\\
		\hline
	\end{tabular}
\end{table}
In Bag mode (Table \ref{BagTable}) DAE performed better than AGPE by 15\%.
\begin{table}[h!]
	\caption{Bag mode}\label{BagTable}
	\centering
	\begin{tabular}{| l || c | c |c |}
		\hline
		Algorithm 			 & $e_{\phi} [deg]$	& $e_{\theta} [deg]$	& $e [deg]$   \\
		\hline\hline
		Madgwick  			 & 0.67 			& 0.62 					& 0.91\\ 
		\hline
		Mahony    			 & 0.64 			& 0.60 					& 0.88\\
		\hline
		ES EKF    			 & 1.03 			& 0.93 					& 1.39\\
		\hline
%		AEKF 1    			 & 2.04 			& 2.75 					& 3.42\\
%		\hline
		AEKF  				 & 0.69 			& 0.68 					& 0.97\\
		\hline
		AGPE		     	 & 0.60 			& $\bm{0.49}$			& 0.77\\
		\hline
		DAE (ours)     		 & $\bm{0.52}$		& $\bm{0.49}$ 		 	& $\bm{0.71}$\\
		\hline
		
	\end{tabular}
\end{table}
Overall performance of all algorithms is summarized in Table \ref{PerformanceSummaryTable}. By average DAE has outperformed AGPE by 10\%. The Mahony, Madwick, and EKF are popular AHRS solutions and frequently referenced in the literature. On average Mahony was the best among them, but not by a great deal (4.2\%). DAE outperformed Mahony by 37\%. The AEKF had the weakest performance among all estimators. In this algorithm the covariance of the linear acceleration signal is estimated and treated as a Gaussian distributed, uncorrelated noise signal. Since the linear acceleration signal does not meet theses definitions, the basic assumptions of the KF framework are violated explaining the filters performance. The performance of the AGPE is quite close to the DAE. In the body mode the AGPE has outperformed the DAE. The DAE and the AGPE share the same basic model. The main difference lies in the training/optimization method. The DAE generalizes from small portions of the training set to full length experiments in test. When optimizing the  AGPE, the whole training data is processed many times in order to estimate the gradient. The consistency in performance is dependent on the distribution of data to train and test sets. Examining the results in \cite{vertzberger2021attitude} we observe that performance degrades from train to test in all modes except on body mode in which accuracy is consistent. This explains the accurate results of AGPE in body mode. These results underline the advantage of the hybrid method over common approaches and shows yet another improvement achieved by the latest learning framework. 
\begin{table}[h!]
	\caption{Performance Summary Table.}\label{PerformanceSummaryTable}
	\centering
	\begin{tabular}{| l || c | c |c |c |c |}
		\hline
		Algorithm 			 & $e_{pocket}$	& $e_{texting}$	& $e_{body}$ & $e_{bag}$	& $e_{average}$  \\
		\hline\hline
		Madgwick  			 & 1.16 		& 1.19 			& 1.46		 & 0.91			& 1.18\\ 
		\hline
		Mahony    			 & 1.51 		& 1.05			& 1.08		 & 0.88			& 1.13\\
		\hline
		ES EKF    			 & 1.11 		& 1.20 			& 1.28	 	 & 1.39			& 1.18\\
		\hline
%		AEKF 1    			 & 2.29 		& 2.27			& 3.01		 & 3.42			& 2.75\\
%		\hline
		AEKF    			 & 1.52 		& 1.21 			& 1.33	 	 & 0.97			& 1.26\\
		\hline
		AGPE		     	 & 0.94 		& 0.86			& $\bm{0.58}$& 0.77			& 0.81\\
		\hline
		DAE (ours)     		 & $\bm{0.68}$	& $\bm{0.85}$  	& 0.68		 & $\bm{0.71}$	& $\bm{0.73}$\\
		\hline
		
	\end{tabular}
\end{table}
%Based on the performance measure defined in  (\ref{PerformanceMeasure}), DAE performs better that AGPE on pocket (Table \ref{PocketTable}) and bag (Table \ref{BagTable}) modes by 28\% and 12\% respectively. On texting (Table \ref{TextingTable}) mode performance is equivalent. On bag mode (Table \ref{BagTable}) AGPE performed better by 15\%..
%
\section{Conclusions}\label{Section_Conclusions}
A hybrid model-learning approach for adaptive attitude estimation was proposed. The initial data driven framework (APGE, AAE), introduced by the authors in \cite{vertzberger2021attitude,vertzberger2021Magnetometers} was elaborated into a hybrid model-learning framework. Instead of using an empiric function obtained by an optimization process, the method integrates a NN nested in a complementary filter based structure. The training process enables stochastic batch optimization and, therefore, performs significantly quicker than the optimization process needed in the former configuration. \\
The NN is designed to predict the optimal fusion weights. Thus, it is carefully optimized to meet the requirements of predicting the accelerometer multiplicative weights. Using the soft threshold custom activation function. Additionally, an input augmentation layer designed to encourage nonlinear behavior, is integrated in the network structure.\\
The performance of our approach has been tested on experimental data compared to a selection of commonly used estimators including Madgwick, Mahony, ES-EKF, and APGE. The DAE approach  performed better than Madgwick, Mahony, and ES-EKF by 37\%. The improvement of the AGPE (both DAE and APGE share a similar model-based structure) averaged 10\%. 

\bibliographystyle{ieeetran}
\bibliography{DeepAHRS_References}

\begin{thebibliography}{10}
\providecommand{\url}[1]{#1}
\csname url@samestyle\endcsname
\providecommand{\newblock}{\relax}
\providecommand{\bibinfo}[2]{#2}
\providecommand{\BIBentrySTDinterwordspacing}{\spaceskip=0pt\relax}
\providecommand{\BIBentryALTinterwordstretchfactor}{4}
\providecommand{\BIBentryALTinterwordspacing}{\spaceskip=\fontdimen2\font plus
\BIBentryALTinterwordstretchfactor\fontdimen3\font minus
  \fontdimen4\font\relax}
\providecommand{\BIBforeignlanguage}[2]{{%
\expandafter\ifx\csname l@#1\endcsname\relax
\typeout{** WARNING: IEEEtran.bst: No hyphenation pattern has been}%
\typeout{** loaded for the language `#1'. Using the pattern for}%
\typeout{** the default language instead.}%
\else
\language=\csname l@#1\endcsname
\fi
#2}}
\providecommand{\BIBdecl}{\relax}
\BIBdecl

\bibitem{titterton2004strapdown}
D.~Titterton, J.~L. Weston, and J.~Weston, \emph{Strapdown inertial navigation
  technology}.\hskip 1em plus 0.5em minus 0.4em\relax IET, 2004, vol.~17.

\bibitem{farrell2008aided}
J.~Farrell, \emph{{Aided navigation: GPS with high rate sensors}}.\hskip 1em
  plus 0.5em minus 0.4em\relax McGraw-Hill, Inc., 2008.

\bibitem{madgwick2011estimation}
S.~O.~H. Madgwick, A.~J.~L. Harrison, and R.~Vaidyanathan, ``{Estimation of IMU
  and MARG orientation using a gradient descent algorithm},'' in \emph{2011
  IEEE international conference on rehabilitation robotics}.\hskip 1em plus
  0.5em minus 0.4em\relax IEEE, 2011, pp. 1--7.

\bibitem{hyde2008estimation}
R.~A. Hyde, L.~P. Ketteringham, S.~A. Neild, and R.~J.~S. Jones, ``Estimation
  of upper-limb orientation based on accelerometer and gyroscope
  measurements,'' \emph{IEEE Transactions on Biomedical Engineering}, vol.~55,
  no.~2, pp. 746--754, 2008.

\bibitem{fourati2010nonlinear}
H.~Fourati, N.~Manamanni, L.~Afilal, and Y.~Handrich, ``{A nonlinear filtering
  approach for the attitude and dynamic body acceleration estimation based on
  inertial and magnetic sensors: Bio-logging application},'' \emph{IEEE Sensors
  Journal}, vol.~11, no.~1, pp. 233--244, 2010.

\bibitem{diaz2014standalone}
E.~M. Diaz, A.~L.~M. Gonzalez, and F.~de~Ponte~M{\"u}ller, ``Standalone
  inertial pocket navigation system,'' in \emph{2014 IEEE/ION Position,
  Location and Navigation Symposium-PLANS 2014}.\hskip 1em plus 0.5em minus
  0.4em\relax IEEE, 2014, pp. 241--251.

\bibitem{diaz2015evaluation}
E.~M. Diaz, F.~de~Ponte~M{\"u}ller, A.~R. Jim{\'e}nez, and F.~Zampella,
  ``{Evaluation of AHRS algorithms for inertial personal localization in
  industrial environments},'' in \emph{2015 IEEE International Conference on
  Industrial Technology (ICIT)}.\hskip 1em plus 0.5em minus 0.4em\relax IEEE,
  2015, pp. 3412--3417.

\bibitem{michel2017attitude}
T.~Michel, P.~Geneves, H.~Fourati, and N.~Laya{\"\i}da, ``On attitude
  estimation with smartphones,'' in \emph{2017 IEEE International Conference on
  Pervasive Computing and Communications (PerCom)}.\hskip 1em plus 0.5em minus
  0.4em\relax IEEE, 2017, pp. 267--275.

\bibitem{renaudin2014magnetic}
V.~Renaudin and C.~Combettes, ``{Magnetic, acceleration fields and gyroscope
  quaternion (MAGYQ)-based attitude estimation with smartphone sensors for
  indoor pedestrian navigation},'' \emph{Sensors}, vol.~14, no.~12, pp.
  22\,864--22\,890, 2014.

\bibitem{mahony2008nonlinear}
R.~Mahony, T.~Hamel, and J.-M. Pflimlin, ``Nonlinear complementary filters on
  the special orthogonal group,'' \emph{IEEE Transactions on automatic
  control}, vol.~53, no.~5, pp. 1203--1218, 2008.

\bibitem{pourtakdoust2007adaptive}
S.~H. Pourtakdoust and H.~G. Asl, ``{An adaptive unscented Kalman filter for
  quaternion-based orientation estimation in low-cost AHRS},'' \emph{Aircraft
  Engineering and Aerospace Technology}, 2007.

\bibitem{de2011uav}
H.~G. De~Marina, F.~J. Pereda, J.~M. Giron-Sierra, and F.~Espinosa, ``{UAV
  attitude estimation using unscented Kalman filter and TRIAD},'' \emph{IEEE
  Transactions on Industrial Electronics}, vol.~59, no.~11, pp. 4465--4474,
  2011.

\bibitem{wang2015quaternion}
L.~Wang, Z.~Zhang, and P.~Sun, ``{Quaternion-based Kalman filter for AHRS using
  an adaptive-step gradient descent algorithm},'' \emph{International Journal
  of Advanced Robotic Systems}, vol.~12, no.~9, p. 131, 2015.

\bibitem{rehbinder2003pose}
H.~Rehbinder and B.~K. Ghosh, ``Pose estimation using line-based dynamic vision
  and inertial sensors,'' \emph{IEEE Transactions on Automatic control},
  vol.~48, no.~2, pp. 186--199, 2003.

\bibitem{lobo2003vision}
J.~Lobo and J.~Dias, ``Vision and inertial sensor cooperation using gravity as
  a vertical reference,'' \emph{IEEE Transactions on Pattern Analysis and
  Machine Intelligence}, vol.~25, no.~12, pp. 1597--1608, 2003.

\bibitem{andersson2019deep}
C.~Andersson, A.~H. Ribeiro, K.~Tiels, N.~Wahlstr{\"o}m, and T.~B. Sch{\"o}n,
  ``Deep convolutional networks in system identification,'' in \emph{2019 IEEE
  58th Conference on Decision and Control (CDC)}.\hskip 1em plus 0.5em minus
  0.4em\relax IEEE, 2019, pp. 3670--3676.

\bibitem{vaswani2017attention}
A.~Vaswani, N.~Shazeer, N.~Parmar, J.~Uszkoreit, L.~Jones, A.~N. Gomez,
  {\L}.~Kaiser, and I.~Polosukhin, ``Attention is all you need,'' in
  \emph{Advances in neural information processing systems}, 2017, pp.
  5998--6008.

\bibitem{asraf2021pdrnet}
O.~Asraf, F.~Shama, and I.~Klein, ``pdrnet: a deep-learning pedestrian dead
  reckoning framework,'' \emph{IEEE Sensors Journal}, 2021.

\bibitem{vertzberger2021attitude}
E.~Vertzberger and I.~Klein, ``Attitude adaptive estimation with smartphone
  classification for pedestrian navigation,'' \emph{IEEE Sensors Journal},
  vol.~21, no.~7, pp. 9341--9348, 2021.

\bibitem{vertzberger2021Magnetometers}
------, ``Attitude and heading adaptive estimation using a data driven
  approach,'' in \emph{International Conference on Indoor Positioning and
  Indoor Navigation (IPIN), 2021}, 2021.

\bibitem{yan2018ridi}
H.~Yan, Q.~Shan, and Y.~Furukawa, ``{RIDI: Robust IMU double integration},'' in
  \emph{Proceedings of the European Conference on Computer Vision (ECCV)},
  2018, pp. 621--636.

\bibitem{black1964passive}
H.~D. Black, ``A passive system for determining the attitude of a satellite,''
  \emph{AIAA Journal}, vol.~2, no.~7, pp. 1350--1351, 1964.

\bibitem{wahba1965least}
G.~Wahba, ``A least square estimate of spacecraft attitude.'' \emph{Siam
  Review, vol. 7, no. 3}, 1965.

\bibitem{shuster1981three}
M.~D. Shuster and S.~D. Oh, ``Three-axis attitude determination from vector
  observations,'' \emph{Journal of Guidance and Control}, vol.~4, no.~1, pp.
  70--77, 1981.

\bibitem{bar1986frequency}
I.~Bar-Itzhack, ``Frequency and time domain designs of a strapdown vertical
  determination system,'' in \emph{Astrodynamics Conference}, 1986, p. 2148.

\bibitem{valenti2015keeping}
R.~G. Valenti, I.~Dryanovski, and J.~Xiao, ``Keeping a good attitude: a
  quaternion-based orientation filter for imus and margs,'' \emph{Sensors},
  vol.~15, no.~8, pp. 19\,302--19\,330, 2015.

\bibitem{sabatini2011kalman}
A.~M. Sabatini, ``{Kalman-filter-based orientation determination using
  inertial/magnetic sensors: Observability analysis and performance
  evaluation},'' \emph{Sensors}, vol.~11, no.~10, pp. 9182--9206, 2011.

\bibitem{munguia2011attitude}
R.~Munguia and A.~Grau, ``{Attitude and heading system based on EKF total state
  configuration},'' in \emph{2011 IEEE International Symposium on Industrial
  Electronics}.\hskip 1em plus 0.5em minus 0.4em\relax IEEE, 2011, pp.
  2147--2152.

\bibitem{park2020adaptive}
S.~Park, J.~Park, and C.~G. Park, ``{Adaptive attitude estimation for low-cost
  MEMS IMU using ellipsoidal method},'' \emph{IEEE Transactions on
  Instrumentation and Measurement}, vol.~69, no.~9, pp. 7082--7091, 2020.

\bibitem{choukroun2006novel}
D.~Choukroun, I.~Y. Bar-Itzhack, and Y.~Oshman, ``{Novel quaternion Kalman
  filter},'' \emph{IEEE Transactions on Aerospace and Electronic Systems},
  vol.~42, no.~1, pp. 174--190, 2006.

\bibitem{lee2012estimation}
J.~K. Lee, E.~J. Park, and S.~N. Robinovitch, ``Estimation of attitude and
  external acceleration using inertial sensor measurement during various
  dynamic conditions,'' \emph{IEEE Transactions on Instrumentation and
  Measurement}, vol.~61, no.~8, pp. 2262--2273, 2012.

\bibitem{feng2017new}
K.~Feng, J.~Li, X.~Zhang, C.~Shen, Y.~Bi, T.~Zheng, and J.~Liu, ``{A new
  quaternion-based Kalman filter for real-time attitude estimation using the
  two-step geometrically-intuitive correction algorithm},'' \emph{Sensors},
  vol.~17, no.~9, p. 2146, 2017.

\bibitem{ding2007improving}
W.~Ding, J.~Wang, C.~Rizos, and D.~Kinlyside, ``{Improving adaptive Kalman
  estimation in GPS/INS integration},'' \emph{The Journal of Navigation},
  vol.~60, no.~3, p. 517, 2007.

\bibitem{wang2004adaptive}
M.~Wang, Y.~Yang, R.~R. Hatch, and Y.~Zhang, ``{Adaptive filter for a miniature
  MEMS based attitude and heading reference system},'' in \emph{PLANS 2004.
  Position Location and Navigation Symposium (IEEE Cat. No. 04CH37556)}.\hskip
  1em plus 0.5em minus 0.4em\relax IEEE, 2004, pp. 193--200.

\bibitem{al2019deep}
M.~K. Al-Sharman, Y.~Zweiri, M.~A.~K. Jaradat, R.~Al-Husari, D.~Gan, and L.~D.
  Seneviratne, ``Deep-learning-based neural network training for state
  estimation enhancement: Application to attitude estimation,'' \emph{IEEE
  Transactions on Instrumentation and Measurement}, vol.~69, no.~1, pp. 24--34,
  2019.

\bibitem{russo2020danae}
P.~Russo, F.~Di~Ciaccio, and S.~Troisi, ``Danae: a denoising autoencoder for
  underwater attitude estimation,'' \emph{arXiv preprint arXiv:2011.06853},
  2020.

\bibitem{russo2021danae++}
------, ``Danae++: a smart approach for denoising underwater attitude
  estimation,'' \emph{Sensors}, vol.~21, no.~4, p. 1526, 2021.

\bibitem{brossard2020denoising}
M.~Brossard, S.~Bonnabel, and A.~Barrau, ``{Denoising IMU Gyroscopes with Deep
  Learning for Open-Loop Attitude Estimation},'' \emph{arXiv preprint
  arXiv:2002.10718}, 2020.

\bibitem{esfahani2019orinet}
M.~A. Esfahani, H.~Wang, K.~Wu, and S.~Yuan, ``{OriNet: robust 3-D orientation
  estimation with a single particular IMU},'' \emph{IEEE Robotics and
  Automation Letters}, vol.~5, no.~2, pp. 399--406, 2019.

\bibitem{weber2005neural}
D.~Weber, C.~G{\"u}hmann, and T.~Seel, ``Neural networks versus conventional
  filters for inertial-sensor-based attitude estimation. arxiv 2020,''
  \emph{arXiv preprint arXiv:2005.06897}.

\bibitem{weber2021riann}
------, ``Riann: a robust neural network outperforms attitude estimation
  filters,'' \emph{arXiv preprint arXiv:2104.07391}, 2021.

\bibitem{chen2018ionet}
C.~Chen, X.~Lu, A.~Markham, and N.~Trigoni, ``{Ionet: Learning to cure the
  curse of drift in inertial odometry},'' \emph{arXiv preprint
  arXiv:1802.02209}, 2018.

\bibitem{yan2019ronin}
H.~Yan, S.~Herath, and Y.~Furukawa, ``{RoNIN: robust neural inertial navigation
  in the wild: benchmark, evaluations, and new methods},'' \emph{arXiv preprint
  arXiv:1905.12853}, 2019.

\bibitem{he2016deep}
K.~He, X.~Zhang, S.~Ren, and J.~Sun, ``Deep residual learning for image
  recognition,'' in \emph{Proceedings of the IEEE conference on computer vision
  and pattern recognition}, 2016, pp. 770--778.

\bibitem{smola2004tutorial}
A.~J. Smola and B.~Sch{\"o}lkopf, ``A tutorial on support vector regression,''
  \emph{Statistics and Computing}, vol.~14, no.~3, pp. 199--222, 2004.

\end{thebibliography}

%\newpage

\begin{IEEEbiography}[{\includegraphics[width=1in,height=1.25in,clip,keepaspectratio]{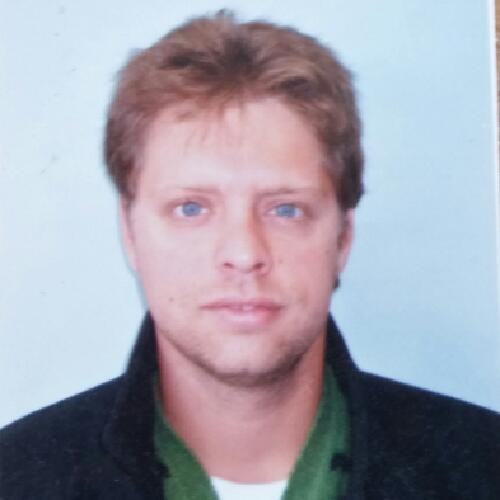}}]{Eran Vertzberger} received the B.Sc. and M.Sc. degrees in mechanical engineering from the Technion - Israel Institute of Technology, Haifa, Israel, in 2009 and 2014, respectively. He is a PhD candidate at the Department of Marine Technologies, University of Haifa, Israel. His research interests include navigation theory, sensor fusion and optimal control for robotic applications.
\end{IEEEbiography}
\begin{IEEEbiography}[{\includegraphics[width=1in,height=1.25in,clip,keepaspectratio]{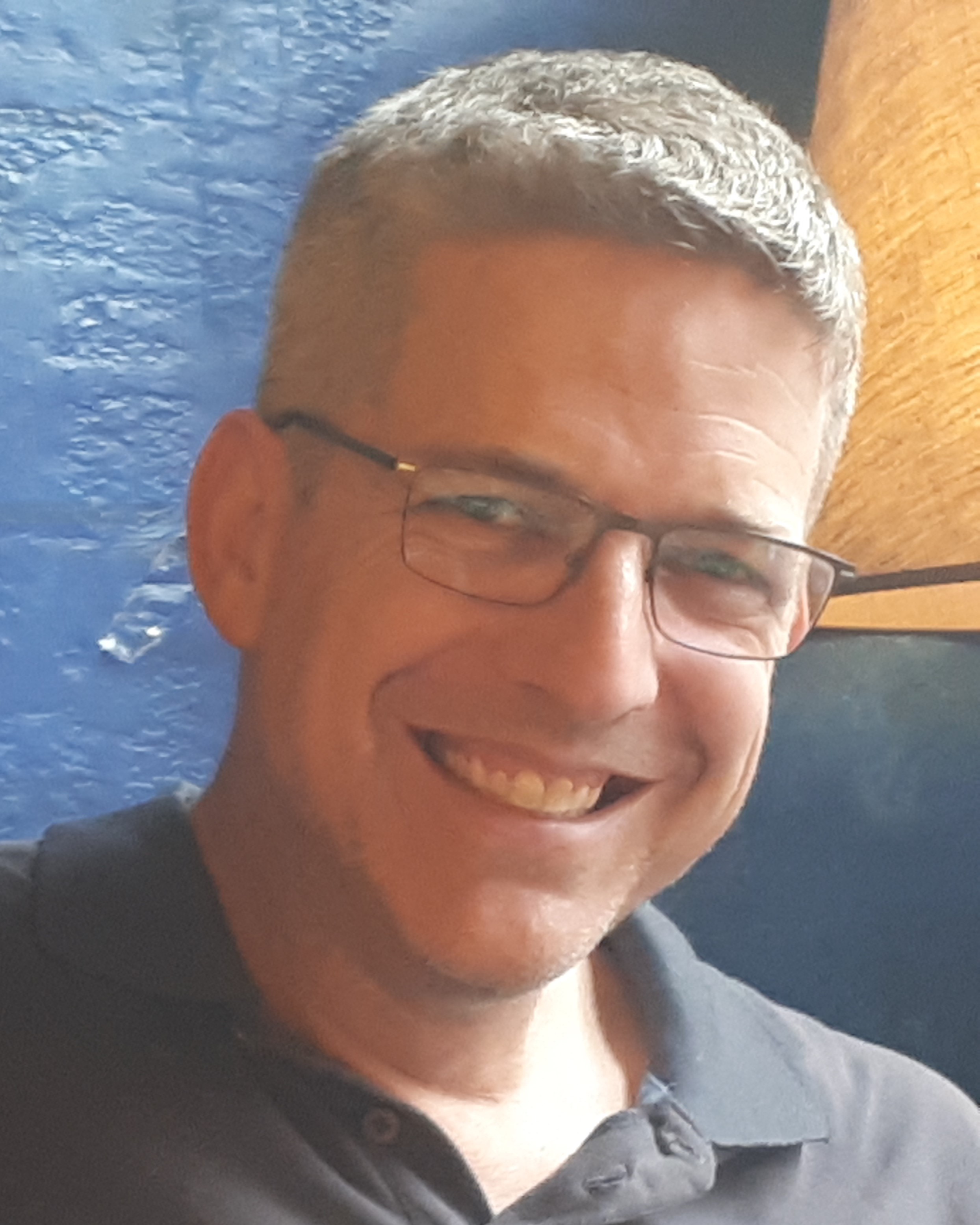}}]{Itzik Klein} (Senior Member, IEEE) received the B.Sc. and M.Sc. degrees in Aerospace Engineering from the Technion - Israel Institute of Technology, Haifa, Israel, in 2004 and 2007, respectively, and a Ph.D. degree in Geo-Information Engineering from the Technion Israel Institute of Technology, in 2011. He is currently an Assistant Professor, heading the Autonomous Navigation and Sensor Fusion Lab, at the Department of Marine Technologies, University of Haifa, Israel. His research interests include data-driven based navigation, novel inertial navigation systems architectures, autonomous underwater vehicles, sensor fusion, and estimation theory.
\end{IEEEbiography}

\vfill

\end{document}